\documentclass[a4paper,11pt]{article}
\pdfoutput=1 

\usepackage{jcappub} 

\usepackage[T1]{fontenc} 
\usepackage[utf8]{inputenc}
\usepackage{amsmath,amssymb}
\usepackage{orcidlink}

\title{\boldmath Cosmological perturbation theory in $f(Q,T)$ gravity}

\author[a]{Antonio Nájera\orcidlink{0000-0001-9738-7704}}

\author[b]{, Amanda Fajardo\orcidlink{0000-0001-7014-4278}}

\affiliation[a]{Instituto de Ciencias Nucleares, Universidad Nacional
	Aut\'onoma de M\'exico, Circuito Exterior C.U., A.P. 70-543,
	Ciudad de M\'exico 04510, M\'exico}
	
\affiliation[b]{Facultad de Ciencias Físico-Matemáticas, Universidad Autónoma de Nuevo León, Av. Pedro de Alba S/N, Ciudad Universitaria
Monterrey, Nuevo León 66455, México}

\emailAdd{antonio.najera@correo.nucleares.unam.mx}
\emailAdd{amanda.fajardogr@uanl.edu.mx}

\abstract{We developed the cosmological linear theory of perturbations for $f(Q,T)$ gravity, which is an extension of symmetric teleparallel gravity, with $Q$ the non-metricity and $T$ the trace of the stress-energy tensor. By considering an ansatz of $f(Q,T)=f_1(Q)+f_2(T)$, which has been broadly studied in the literature and the coincident gauge where the connection vanishes, we got equations consistent with $f(Q)$ gravity when $f_{T}=0$. In the case of the tensor perturbations, the propagation of gravitational waves was found to be identical to $f(Q)$, as expected. For scalar perturbations, outside the limit $f_T = 0$, we got that the coupling between $Q$ and $T$ in the Lagrangian produces a coupling between the perturbation of the density and the pressure. This coupling is preserved when considering the weak coupling limit between $Q$ and $T$. On the other hand, in the strong coupling limit with a generic function of the form $f_2(T) = \alpha T + \beta T^2$, the perturbative equations are heavily driven by the $f_2(T)$ derivatives when $\beta \neq 0$. However, when $\beta=0$, the perturbative equations are identical to the weak coupling limit even though this case is a non-minimally coupling one. The presence of $T$ in the Lagrangian breaks the equation of the conservation of energy, which in turn breaks the standard $\rho' + 3\mathcal{H} (\rho+p) = 0$ relation. We also derived a coupled system of differential equations between $\delta$, the density contrast and $v$ in the $\mathcal{H}<<k$ limit and with negligible time derivative of the scalar perturbation potentials, which will be useful in future studies to see whether this class of theories constitute a good alternative to dark matter. These results might also enable to test $f(Q,T)$ gravity with CMB and standard siren data that will help to determine if these models can reduce the Hubble constant tension and if they can constitute an alternative to the $\Lambda$CDM model.}

\begin{document}
\maketitle
\flushbottom

\section{Introduction}
\label{sec:introduction}

In General Relativity (GR) and extensions of it, gravity is caused by curvature in spacetime with the aid of the metric and the Levi-Civita connection. However, GR works on a pseudo Riemannian spacetime and hence, if a simpler formulation of gravity is wanted, we need to consider alternatives. A possible form to do this is to work on a spacetime with zero curvature by setting the Riemann tensor to zero ($R^\alpha_{\;\;\beta\mu\nu} = 0$). This is known as the teleparallel formulation \cite{maluf2013teleparallel,bahamonde2021teleparallel} where is possible to build the Weitzenböck connection $\Gamma^\alpha_{\;\;\mu\nu} = e^{a\alpha} \partial_\mu e_{a\nu}$ \cite{maluf2013teleparallel}. If we impose the additional constraint $\nabla_\alpha g_{\mu\nu}=0$, i.e., a metric compatible connection, gravity will be driven by the torsion tensor $T^\alpha_{\;\;\mu\nu} = 2\Gamma^\alpha_{\;\;[\mu\nu]}$. This class of theories are called metric teleparallel gravity \cite{bahamonde2021teleparallel} and in this framework, the connection is not symmetric in its lower indices. However, if we work in a torsion-less space, gravity is driven by the non-metricity which is defined as $Q_{\alpha\mu\nu} \equiv \nabla_\alpha g_{\mu \nu}$. This final class of theories are called symmetric teleparallel gravity \cite{Nester:1998mp,Adak:2005cd} and in they, there is a gauge where the connection vanishes, called the coincident gauge \cite{beltran2019geometrical}. Furthermore, in both, metric teleparallel and symmetric teleparallel a GR equivalent can be built, which has the same equations as GR. They are called the teleparallel equivalent of GR \cite{maluf2013teleparallel} and the symmetric teleparallel equivalent of GR \cite{jimenez2020cosmology,jimenez2018coincident}, respectively. \\

In the present paper, we focused in the third form to describe gravity, where the non-metricity drives the gravitational force with zero torsion and curvature. In order to get the symmetric teleparallel equivalent of GR (STEGR), we need to assume a specific form of the action ($S = \int \sqrt{-g} (\kappa^{-2}Q + \mathcal{L}_m)$) with $\kappa^2= 8\pi G$ and $Q$ the non-metricity scalar, a scalar given in terms of contractions of the non-metricity tensor. Because of this, this scalar is analogue to the Ricci scalar in GR.\\

With this new formulation of gravity, we have solved the problem of working on a pseudo Riemannian spacetime. However, since we have the STEGR, we still carry several modern problems in cosmology. For example, if we combine the STEGR with the $\Lambda$CDM model, we still have the cosmological constant problem. Therefore, it is useful to consider extensions within the symmetric teleparallel gravity framework. In a similar form as in $f(R)$ gravity, we can consider an extension by making the generalisation $Q \to f(Q)$. This class of theories were proposed in \cite{jimenez2018coincident} and they give generalised Friedmann equations, which can potentially explain the accelerated expansion of the Universe in terms of geometry instead of a cosmological constant. In addition to this, we can consider another extension to these $f(Q)$ theories. By coupling the trace of the stress-energy tensor with the non-metricity scalar in the Lagrangian, $f(Q,T)$ gravity is formulated \cite{xu2019f}. Theories with a coupling between geometry and matter are particularly interesting since the stress-energy conservation is broken \cite{xu2019f,harko2011f}. This has astonishing consequences, stress-energy transfer between matter and geometry and particle production/annihilation \cite{xu2019f,wu2018palatini}, which might have a connection with quantum gravity. \\

Until now, several functions of $f(Q,T)$ have been tested in a background framework, for example \cite{Arora:2020iva,Arora:2020tuk,Arora:2020met}. Moreover, they seem promising to challenge the standard $\Lambda$CDM model because the function $f(Q,T) = -(Q+2\Lambda)/G_N - ((16\pi)^2 G_N b)/(120 H_0^2) T^2$ showed a substantial preference against $\Lambda$CDM using Cosmic Chronometers and SNeIa data \cite{najera2021fitting}. This motivates studies of $f(Q,T)$ outside of the background framework. In this paper, we develop the theory of cosmological linear perturbations in $f(Q,T)$ gravity, which will enable us to perform future tests that will determine if these theories can constitute an alternative to $\Lambda$CDM. These tests will include testing $f(Q,T)$ gravity with CMB data such as Planck \cite{Planck:2018vyg}. Moreover, by getting the equation of the density contrast, it will be possible to see if this theories constitute an alternative to dark matter. The study of this alternative models is encouraged by the problems that the standard $\Lambda$CDM presents such as the Hubble constant and the cosmological constant problems. \\

The standard model of cosmology $\Lambda$CDM produces an accurate description of numerous observational values in astrophysics and astronomy. For example, this model provides an excellent fit to the anisotropies observed in the CMB \cite{Planck:2018vyg}, it forecasts correctly observables such as the cosmic expansion rate, and predicts a distribution of large-scale structures that agrees with observations \cite{Peebles:2002gy}. Nevertheless, it is still incapable to provide a concrete definition of central concepts in the theory such as the nature of cold dark matter and the cosmological constant. Furthermore, there has been an statistical discrepancy of more than $4 \sigma$ between the value of the Hubble constant $H_0$ of the SH0ES collaboration \cite{Riess:2019cxk} and the value estimated by the Planck Collaboration \cite{Planck:2018vyg}. In addition to this, the Pantheon SNeIa sample \cite{Pan-STARRS1:2017jku,krishnan2020there,krishnan2021running} predicts a decreasing trend in the $H_0$ value with increasing the redshift when taking bins of the sample which can be due to modified gravity models or systematic errors \cite{Dainotti:2021pqg,DeSimone:2021rus}. \\

This motivates a search for systematic errors in both experiments and theoretical alternatives to the standard model, looking to reconcile the values of the modern value of the Hubble constant $H_0$ for low and high-redshift probes. A great number of different approaches (See \cite{divalentino2021} for a review) tried to reconcile the $H_0$ tension. One example of this is modified gravity which is the approach we take in this paper, in particular linear cosmological perturbation theory in $f(Q,T)$ gravity. \\

The scalar-vector-tensor (SVT) decomposition is an essential resource to explore first order perturbations. In this theory, the perturbed metric can be written in terms of its longitudinal part plus its orthogonal part plus its transverse part \cite{piattella2018lecture}. In total, these three classes are made up of 4 scalars $\Psi$, $W$,$\Phi$ and $E$, two vectors $W_i$ and $V_i$, and a tensor $h_{ij}^T$ that is the transverse contribution and which constitutes the tensor perturbation \cite{piattella2018lecture}. In first order perturbations, the three classes do not mix up, so they can be analysed separately. \\

In this paper, we analysed the cosmological perturbations of $f(Q,T)$ theories of the form $f(Q,T) = f_1(Q) + f_2(T)$ since this class of functions has been largely studied in the literature \cite{gadbail2021viscous,Agrawal:2021rur,Gadbail:2021kgd,Pati:2021ach,Pradhan:2021dpf,Arora:2021tuh,Godani:2021mld,Arora:2021jik,Bhattacharjee:2021hwm,Zia:2021vhr,Arora:2020iva,Arora:2020tuk,Arora:2020met,najera2021fitting}. In order to do this, we worked with the conformal time taken as the zeroth coordinate and the background metric described as an additive function of the Friedmann–Lemaître–Robertson–Walker metric $\bar{g}_{\mu\nu}$ and a first order perturbation $h_{\mu\nu}$. In section \ref{sec:fQTtheory} we present the general theory of $f(Q,T)$ gravity needed to develop perturbation theory. Using the SVT decomposition on the metric, we found the tensor, vector and scalar perturbations of the field equations in section \ref{sec:cosmoPerturbations}. In section \ref{sec:densityConstrast}, we analysed the density contrast equation that comes from scalar perturbations which will make possible to test if $f(Q,T)$ gravity represents a viable alternative to dark matter. In order to explore the effect of the coupling of $T$ in the Lagrangian, we studied the weak coupling limit between geometry and matter in section \ref{sec:weakCoupling} and the strong coupling limit in section \ref{sec:strongCoupling}. Finally, in section \ref{sec:conclusions} we present a summary of the main results and findings.

\section{$f(Q,T)$ Theory}
\label{sec:fQTtheory}
The general action for $f(Q,T)$ gravity is given by \cite{xu2019f}
\begin{equation}
	\label{eqn:f(Q,T)action}
	S = \int d^4 x \sqrt{-g} \left( \frac{1}{16\pi} f(Q,T) + \mathcal{L}_m \right),
\end{equation}
where $Q$ stands for the non-metricity scalar, $T$ for the trace of the stress-energy tensor, $\mathcal{L}_m$ for the matter Lagrangian and $g$ for the determinant of the metric. Moreover, the non-metricity scalar is defined as \cite{xu2019f}
\begin{equation}
	\label{eqn:nonMetricityScalar}
	Q \equiv - g^{\mu \nu} \left( L^\alpha_{\;\, \beta \mu} L^\beta_{\; \, \nu \alpha} - L^\alpha_{\;\, \beta \alpha} L^\beta_{\; \, \mu \nu} \right), 
\end{equation}
where $L^\alpha_{\;\, \beta \mu}$ is the deformation tensor given by \cite{xu2019f}
\begin{equation}
	L^\alpha_{\;\, \beta \mu} = - \frac{1}{2} g^{\alpha \lambda} \left( \nabla_\mu g_{\beta \lambda} + \nabla_\beta g_{\lambda \mu} - \nabla_\lambda g_{\mu \beta} \right).
\end{equation}

The variation of the action (\ref{eqn:f(Q,T)action}) gives the field equations \cite{xu2019f}
\begin{equation}
	\label{eqn:fieldEquations}
	-\frac{2}{\sqrt{-g}} \nabla_\alpha \left( f_Q \sqrt{-g} P^{\alpha}_{\;\; \mu \nu} \right) - \frac{1}{2} f g_{\mu \nu} + f_T (T_{\mu \nu} + \Theta_{\mu \nu}) - f_Q (P_{\mu \alpha \beta} Q^{\;\; \alpha \beta}_\nu - 2Q^{\alpha \beta}_{\;\;\;\;\mu} P_{\alpha \beta \nu}) = 8\pi T_{\mu \nu},
\end{equation}
where $Q_{\mu \nu \alpha} \equiv \nabla_\mu g_{\nu \alpha}$ is the non-metricity tensor, $f_Q \equiv \dfrac{\partial f}{\partial Q}$, $f_T \equiv \dfrac{\partial f}{\partial T}$, $T_{\mu \nu} = -\dfrac{2}{\sqrt{-g}} \dfrac{\delta(\sqrt{-g} \mathcal{L}_m)}{\delta g^{\mu \nu}}$ the stress-energy tensor, $\Theta_{\mu \nu} = g^{\alpha \beta} \dfrac{\delta T_{\alpha \beta}}{\delta g^{\mu \nu}}$ and $P^\alpha_{\;\;\mu \nu}$ the super potential given by \cite{xu2019f}
\begin{equation}
	\label{eqn:superpotential}
	P^\alpha_{\;\;\mu \nu} = - \frac{1}{2} L^\alpha_{\;\, \mu \nu} + \frac{1}{4} \left( Q^\alpha - \tilde{Q}^\alpha \right) g_{\mu \nu} - \frac{1}{4} \delta^\alpha_{\;\; (\mu }Q_{\nu )},
\end{equation}
where $Q_\alpha \equiv Q_{\alpha \;\, \beta}^{\;\;\,\beta}$ and $\tilde{Q}_\alpha = Q_{\;\; \alpha \beta}^\beta$. In terms of this super potential, the non-metricity scalar is given by $Q = - Q_{\alpha \beta \gamma} P^{\alpha \beta \gamma}$. If we instead vary the action with respect to the connection, with the Lagrange multiplier method, and the conditions $T^\alpha_{\;\;\mu\nu} = 0 = R^\alpha_{\;\;\beta\mu\nu}$ we get the connection field-equations \cite{xu2019f}
\begin{equation}
    \label{eqn:connectionFieldEquations}
    \nabla_\mu \nabla_\nu \left( \sqrt{-g} f_Q P^{\mu\nu}_{\;\;\;\;\alpha} + 4\pi H_\alpha^{\;\;\mu\nu} \right) = 0,
\end{equation}
where $H_\alpha^{\;\;\mu\nu}$ is called hypermomentum and it is given by \cite{xu2019f}
\begin{equation}
    \label{eqn:hypermomentum}
    H_\alpha^{\;\;\mu\nu} = \frac{\sqrt{-g}}{16\pi} f_T \frac{\delta T}{\delta \Gamma^{\alpha}_{\;\;\mu\nu}} + \frac{\delta \sqrt{-g} \mathcal{L}_m}{\delta \Gamma^{\alpha}_{\;\;\mu\nu}},
\end{equation}
with $\Gamma^{\alpha}_{\;\;\mu\nu}$ the connection. Since we will deal with the crossed $T^\mu_{\;\;\nu}$ stress-energy tensor for simplicity, it is useful to derive the field equations (\ref{eqn:fieldEquations}) with an index raised
\begin{equation}
	\label{eqn:fieldEquationsIndexRaised}
	-\frac{2}{\sqrt{-g}} \nabla_\alpha (f_Q \sqrt{-g} P^{\alpha \mu}_{\;\;\;\;\; \nu}) - \frac{1}{2} \delta^\mu_{\;\;\nu} f + f_T (T^\mu_{\;\;\, \nu} + \Theta^\mu_{\;\;\, \nu}) - f_Q P^{\mu \alpha \beta} Q_{\nu \alpha \beta} = 8\pi T^\mu_{\;\;\, \nu}.
\end{equation}

In order to compute the field equations, we need to define a connection which will tell us how the covariant derivatives are written. $f(Q,T)$ theory is an extension of symmetric teleparallel gravity theory where curvature and torsion are both equal to zero and the non-metricity is not trivial and drives the gravitational field. These assumptions can be written as
\begin{equation}
\label{eqn:zeroRiemman}
	R^\alpha_{\;\; \beta\mu \nu} = 0,
\end{equation}
\begin{equation}
\label{eqn:zeroTorsion}
	T^\mu_{\;\;\alpha \beta} = 0,
\end{equation}
\begin{equation}
\label{eqn:nonMetricity}
	\nabla_\alpha g_{\mu \nu} \neq 0.
\end{equation}

The first condition (\ref{eqn:zeroRiemman}) makes the connection integrable and because of that, the connection can be written as \cite{beltran2019geometrical,hohmann2021variational}
\begin{equation}
    \Gamma^\mu_{\;\;\nu\beta} = \left( \Lambda^{-1} \right)^\mu_{\;\;\gamma} \partial_\beta \Lambda^\gamma_{\;\;\nu},
\end{equation}
which gives a vanishing Riemann tensor \cite{hohmann2021variational}. In addition to this, the second condition (\ref{eqn:zeroTorsion}) guarantees that the connection is symmetric in its lower indexes $\left(\Gamma^\mu_{\;\;[\nu\beta]}=0\right)$. This fact causes the constraint $\partial_{[\beta} \Lambda^\gamma_{\;\;\nu]} = 0$. Therefore, these Lambda tensors can be written in terms of a set of parameters $\xi^\gamma$ \cite{beltran2019geometrical}
\begin{equation}
    \Gamma^\gamma_{\;\;\nu} = \partial_\nu \xi^\gamma.
\end{equation}

Therefore, the connection is given by \cite{runkla2018family,beltran2019geometrical}
\begin{equation}
	\label{eqn:generalConnection}
	\Gamma^\mu_{\;\;\nu \alpha} = \frac{\partial x^\mu}{\partial \xi^\beta} \partial_\nu \partial_\alpha \xi^\beta,
\end{equation}
If we let the parameters $\xi^\beta$ be the coordinates ($\xi^\beta = x^\beta$), the connection vanishes. The gauge in which we set these parameters as the coordinates is called coincident gauge and it will be the one in which we will work in this paper. This gauge can be interpreted as the gauge where the origin of the tangent space parameterised by $\xi^\beta$ \textit{coincides} with the origin of the spacetime \cite{beltran2019geometrical}.
However, by taking this gauge we lose the diffeomorphism invariance of the equations. Then, if we want to see how the field equations behave in other frames we need to compute the gauge transformations of the equations as in \cite{jimenez2020cosmology}. In this gauge, the non-metricity tensor is
\begin{equation}
	Q_{\alpha \mu \nu} = \partial_\alpha g_{\mu \nu}.
\end{equation}

Using the definition of $T_{\mu \nu}$, the $\Theta_{\mu \nu}$ can be written as
\begin{equation}
\Theta_{\mu \nu} = \mathcal{L}_m g_{\mu \nu} - 2 T_{\mu \nu}.
\end{equation}

We will consider that the Universe is composed of a perfect fluid. Therefore, the stress-energy tensor is given by
\begin{equation}
\label{eqn:stressEnergy}
    T^\mu_{\;\;\nu} = (\rho + p) u^\mu u_\nu + p \delta^\mu_\nu,
\end{equation}
where $\rho$ is the density and $p$ the pressure. Moreover, we will take the matter Lagrangian as $\mathcal{L}_m = p$ and hence
\begin{equation}
    \Theta^\mu_{\;\;\nu} = p \delta^\mu_\nu - 2 T^\mu_{\;\;\nu}.
\end{equation}

However, there are alternate forms of the matter Lagrangian, for example $\mathcal{L}_m = -\rho$. Since the tensor $\Theta_{\mu\nu}$ is defined in terms of it, by choosing an alternate form of this Lagrangian, the field equations (\ref{eqn:fieldEquations}) would be different. Furthermore, the perturbation equations that will be computed in this paper would differ. Hence, we need to point out that the results presented in this paper assume a specific form of the matter Lagrangian given by $\mathcal{L}_m = p$, as it is standard in $f(Q,T)$ gravity \cite{xu2019f,gadbail2021viscous,Agrawal:2021rur,Gadbail:2021kgd,Pati:2021ach,Pradhan:2021dpf,Arora:2021tuh,Godani:2021mld,Arora:2021jik,Bhattacharjee:2021hwm,Zia:2021vhr,Arora:2020iva,Arora:2020tuk,Arora:2020met,najera2021fitting}.

\subsection{Stress-energy balance equation}

In the case of $f(Q,T)$ gravity, the stress-energy conservation is not satisfied. It is instead given by \cite{xu2019f}
\begin{align}
    \mathcal{D}_\mu T^\mu_{\;\;\nu} &= \frac{1}{f_T - 8\pi} \biggl[ -\mathcal{D}_\mu (f_T \Theta^\mu_{\;\;\nu}) - \frac{16\pi}{\sqrt{-g}} \nabla_\alpha \nabla_\mu H_{\nu}^{\;\;\alpha \mu} + 8\pi \nabla_\mu \left( \frac{1}{\sqrt{-g}} \nabla_\alpha H_\nu^{\;\;\alpha\mu} \right) \nonumber\\
    &- 2 \nabla_\mu A^\mu_{\;\;\nu} + \frac{1}{2} f_T \partial_\nu T \biggr],
\end{align}
where $\sqrt{-g} A^\nu_{\;\;\alpha} = \nabla_\mu \left(\sqrt{-g} f_Q P^{\mu\nu}_{\;\;\;\;\alpha} + 4\pi H_\alpha^{\;\;\mu\nu}\right)$ and $\mathcal{D}_\mu$ represents a covariant derivative with respect to the Levi-Civita connection. With the aid of the connection field equations (\ref{eqn:connectionFieldEquations}), we can rewrite this as
\begin{align}
    \label{eqn:energyBalanceEquiation}
   - (f_T + 8\pi) \mathcal{D}_\mu T^\mu_{\;\; \nu} &+ f_T \partial_\nu p - \frac{1}{2} f_T \partial_\nu T = \frac{1}{\sqrt{-g}} Q_\mu \nabla_\alpha (\sqrt{-g} f_Q P^{\alpha\mu}_{\;\;\;\;\nu}) \nonumber \\
   &+ \frac{2}{\sqrt{-g}} \nabla_\mu \nabla_\alpha (\sqrt{-g} f_Q P^{\mu\alpha}_{\;\;\;\;\nu}) \equiv B_\nu. 
\end{align}

We can now compute the equation for $\rho'$ (the 0 component)
\begin{equation}
\label{eqn:rhoPrimeBack}
    \rho' = -\frac{3\mathcal{H} (f_T + 8\pi) \rho (1 + w)}{f_T + 8\pi + \frac{1}{2} f_T (1-c_s^2)},
\end{equation}
where $c_s^2 = \dfrac{p'}{\rho'}$ and $w = \dfrac{p}{\rho}$ and the prime denotes derivative with respect to the conformal time $\eta$. Notice that we are working with the conformal time since it will be easier to use it when doing perturbation theory. In the limit $f_T=0$, equation (\ref{eqn:rhoPrimeBack}) reduces to the standard one ($\rho' + 3\mathcal{H} \rho (1+w) = 0$ ) and hence the presence of the trace of the stress-energy tensor in the Lagrangian breaks the energy conservation. To see the physical interpretation of this equation, we can write this expression as
\begin{equation}
    \rho' + 3\mathcal{H} \rho(1+w) = - \frac{(3-c_s^2)\rho' + 6\mathcal{H}\rho(1+w)}{16\pi} f_T,
\end{equation}
and the right-hand side of this equation corresponds to the deviation from the standard case, which disappears if $f_T = 0$. This term is therefore a source term that accounts for the stress-energy transfer between matter and geometry and particle production/annihilation \cite{xu2019f,wu2018palatini}.

\section{Cosmological Perturbations}
\label{sec:cosmoPerturbations}

We will study linear cosmological perturbation theory in the framework of $f(Q,T)$ gravity and it is instructive to compare the results with $f(Q)$ gravity \cite{jimenez2020cosmology} since $f(Q,T)$ is an extension of this theory. We will work with conformal time as the zeroth coordinate because by doing it, the background metric can be written as
\begin{equation}
    g_{\mu\nu} = a^2 (\eta_{\mu\nu} + h_{\mu\nu}),
\end{equation}
with $\eta_{\mu\nu} = diag(-1,1,1,1)$ the Minkowski flat metric and $h_{\mu\nu}$ a first order perturbation. The most general form of the perturbation tensor in the scalar-vector-tensor decomposition (SVT) is given by \cite{MUKHANOV1992203,piattella2018lecture}
\begin{equation}
\label{eqn:metricSVT}
h_{\mu \nu} = \begin{pmatrix}
-2\psi & \partial_i W + W_i \\
\partial_i W + W_i & \qquad \qquad 2\left[-\phi \delta_{ij} + \left(\partial_i \partial_j - \frac{1}{3} \delta_{ij} \partial_k \partial_k \right) E+ \partial_{(i} E_{j)}\right] + h^{TT}_{ij} 
\end{pmatrix},
\end{equation}
where the vectorial parts are divergenless, i.e., $\partial_i W_i = 0 = \partial_i E_i = 0$ and the tensor part is traceless and divergenless, i.e., $\partial_i h^T_{ij} = 0 = h^T_{ii}$. It is convenient to define a $\Phi$ scalar given by \cite{jimenez2020cosmology}
\begin{equation}
    \Phi = \phi + \frac{1}{3} \partial_k \partial_k E,
\end{equation}
if we introduce it in the form of $h_{\mu \nu}$ we get
\begin{equation}
\label{eqn:perturbation}
h_{\mu \nu} = \begin{pmatrix}
-2\psi & \partial_i W + W_i \\
\partial_i W + W_i & \qquad \qquad 2\left[-\Phi \delta_{ij} + \partial_i \partial_j  E+ \partial_{(i} E_{j)}\right] + h^{TT}_{ij} \quad
\end{pmatrix},
\end{equation}
by using this form of $h_{\mu \nu}$, the equations of the scalar perturbations will be simplified. Moreover, by performing a perturbation on the mixed stress energy tensor (\ref{eqn:stressEnergy})  \cite{piattella2018lecture}
\begin{equation}
\label{eqn:perturbedStressEnergy}
T^\mu_{\;\;\,\nu} = \begin{pmatrix}
-\bar{\rho} - \delta \rho & \qquad (\bar{\rho} + \bar{p}) (\partial_i v + v_i) \quad \\
-(\bar{\rho} + \bar{p}) (\partial_i (v - W) + (v_i - W_i)) & \qquad (\bar{p}+\delta p) \delta_{ij} 
\end{pmatrix},
\end{equation}
as with $h_{\mu \nu}$, the vector parts are divergenless. The bar over the density and pressure means the density and pressure at background and we will use this notation throughout the paper to distinguish the background and perturbative quantities. Note that we are not taking into account the anisotropic stress $\Pi_{ij}$. This is because if we included, we would need to find a matter Lagrangian that takes into account the presence of this tensor and $\mathcal{L}_m = p$ just takes into account the stress-energy tensor given by equation (\ref{eqn:stressEnergy}). We will start with the tensor perturbations, then we will continue with the vector ones and we will finish with the scalar ones. The perturbation computations will be done with the \texttt{Pytearcat} Python tensor algebra calculator \cite{martin2021pytearcat}. 

Before we start giving the equations, we will state how the perturbations of the functions $f$, $f_T$ and $f_Q$ are written. Since $f(Q,T)$ is a function of both the non-metricity and the trace of the stress-energy tensor, its variation is given by
\begin{equation}
    \delta f = \bar{f}_Q \delta Q + \bar{f}_T \delta T,
\end{equation}
where $\delta Q$ and $\delta T = - \delta \rho + 3 \delta p$ are the perturbation of the non-metricity scalar and the perturbation of the trace of the stress-energy tensor respectively. The variation of $f_T$ and $f_Q$ should have similar expressions, however in this paper, we will work with $f(Q,T)$ functions of the form
\begin{equation}
    f(Q,T) = f_1(Q) + f_2(T),
\end{equation}
i.e., that the function does not have mixed terms. We will take this form of the $f(Q,T)$ because it simplifies considerably the perturbative equations and because this form of the function has been studied broadly in the literature \cite{gadbail2021viscous,Agrawal:2021rur,Gadbail:2021kgd,Pati:2021ach,Pradhan:2021dpf,Arora:2021tuh,Godani:2021mld,Arora:2021jik,Bhattacharjee:2021hwm,Zia:2021vhr,Arora:2020iva,Arora:2020tuk,Arora:2020met,najera2021fitting}. Hence the variation of $f_Q$ and $f_T$ are given by
\begin{equation}
    \delta f_Q = \bar{f}_{QQ} \delta Q,
\end{equation}
and
\begin{equation}
    \delta f_T = \bar{f}_{TT} \delta T. 
\end{equation}

\subsection{Tensor Perturbations}

Let us focus on the tensorial perturbations first. Then, the metric is given by
\begin{equation}
g_{0 0} = -a^2 \quad , g_{0i} = 0 \quad g_{ij} = a^2 (\delta_{ij} + h^{TT}_{ij}),
\end{equation}
with $h^T_{ij}$ divergenless and traceless. Perturbing the field equations (\ref{eqn:fieldEquationsIndexRaised}) and taking the tensor part of the perturbed stress-energy tensor (identical to 0 because we are not taking into account the anisotropic stress)
\begin{equation}
    h''_{ij} + \left( 2\mathcal{H} - (\log f_Q)' \right) h'_{ij} + k^2 h_{ij} = 0.
\end{equation}

This equation describes the evolution of gravitational waves in $f(Q,T)$ and it is identical to the one in $f(Q)$ \cite{jimenez2020cosmology}. This was expected since at vacuum $T = 0$ and hence $f(Q,0) = f(Q)$. Furthermore, this expression enables us to study this class of models with standard sirens. Because of the additional term of $h'_{ij}$, the standard candle luminosity distance $d_L^{\text{em}}(z)$ differs from the gravitational wave luminosity distance $d_L^{\text{gw}}(z)$. This is due to the fact that this additional friction term causes the amplitude of GWs to decay as $|h_{ij}| \propto 1/(a \sqrt{f_Q})$ \cite{jimenez2020cosmology}. If we write the more general propagation equation of the form \cite{belgacem2018gravitational}
\begin{equation}
\label{eqn:generalTensorEquations}
    h''_{ij} + 2\mathcal{H} \left( 1 - \delta(\eta) \right) h'_{ij} + k^2 h_{ij} = 0,
\end{equation}
where $\delta(\eta)$ is an arbitrary function of the conformal time. The additional friction term causes the standard luminosity distance and the gravitational wave luminosity distance to differ and to be related by \cite{belgacem2018gravitational}
\begin{equation}
    d_L^{\text{gw}}(z) = d_L^{\text{em}}(z) \exp \left( - \int_0^z dz' \frac{\delta(z')}{1+z'} \right).
\end{equation}

Therefore, by measuring the gravitational wave luminosity distance to a set of standard sirens or their ratio $d_L^{\text{gw}}/d_L^{\text{em}}$, we will be able to study modified gravity models with propagation equations of the form (\ref{eqn:generalTensorEquations}). In particular, we will test these $f(Q,T)$ theories because the delta function in this case is given by
\begin{equation}
    \delta(\eta) = \frac{(\log f_Q)'}{2\mathcal{H}}.
\end{equation}

Therefore, we can constrain $f(Q,T)$ gravity with the aid of standard sirens.

\subsection{Vector Perturbations}

By taking the vector part of the metric now, we get
\begin{equation}
    g_{00} = 0 \quad , \quad g_{0i} = W_i \quad , \quad g_{i0} = 2\partial_{(i} E_{j)},
\end{equation}
with $\partial_i E_i = 0$. By perturbing the field equations (\ref{eqn:fieldEquationsIndexRaised}) and taking the 0i component of the vector part of the perturbed stress-energy tensor we get the vector equation for $v_i$
\begin{equation}
    v_i = - \frac{\bar{f}_Q k^2}{2(\bar{f}_T+8\pi)(\bar{\rho}+\bar{p})a^2} (E'_i - W_i).
\end{equation}

Note that $W_i - E'_i \equiv B_i$ and $B_i$ is gauge invariant \cite{piattella2018lecture}. Hence, this equation is written in terms of a gauge invariant field.

\subsection{Scalar Perturbations}

By perturbing the field equations (\ref{eqn:fieldEquationsIndexRaised}) with the scalar part of $h_{\mu \nu}$ and taking the $00$ component
\begin{align}
\label{eqn:density}
& a^2 \delta \rho \left( 8\pi + \frac{3}{2} \bar{f}_T - \bar{f}_{TT} (\bar{\rho} + \bar{p}) \right) + a^2 \delta p \left( 3\bar{f}_{TT} (\bar{\rho} + \bar{p}) - \frac{1}{2} \bar{f}_T \right) \nonumber \\
&= 6\left( \bar{f}_Q + 12 \frac{\bar{f}_{QQ}}{a^2} \mathcal{H}^2 \right) \mathcal{H} \left(\mathcal{H}\Psi + {\Phi}^{'}\right) + 2\bar{f}_Q k^2 \Phi -2\left( \bar{f}_Q + 3 \frac{\bar{f}_{QQ}}{a^2} \left( {\mathcal{H}}^{'} + \mathcal{H}^2 \right) \right) \mathcal{H} k^2 W.
\end{align}

Now, taking the perturbation of the spacial part of the stress-energy tensor
\begin{equation}
\delta T^{i}_{\;\;j} = \delta p \, \delta_{ij} ,
\end{equation}
by taking the trace and dividing by three
\begin{equation}
\delta p = \frac{1}{3} \delta T^{i}_{\;\;i},
\end{equation}
hence, by taking the trace we can compute the equation for $\delta p$
\begin{align}
\label{eqn:pressure}
& \frac{1}{4} a^2 \delta \rho \bar{f}_T - a^2 \left( 4\pi + \dfrac{3}{4} \bar{f}_T \right) \delta p = \left( \bar{f}_Q + 12 \frac{\bar{f}_{QQ}}{a^2} \mathcal{H}^2 \right) \left( \mathcal{H} {\Psi}^{'} + {\phi}^{''} \right) + \biggl( \bar{f}_Q \left( {\mathcal{H}}^{'} + 2\mathcal{H}^2 - \dfrac{1}{3} k^2 \right) \nonumber \\
&+ 12 \frac{\bar{f}_{QQ}}{a^2} \mathcal{H}^2 \left(4 {\mathcal{H}}^{'} - \mathcal{H}^2 \right) + 12 \frac{{\bar{f}_{QQ}}^{\,'}}{a^2} \mathcal{H}^3 \biggr) \Psi + 2 \biggl( \bar{f}_Q + 6 \frac{\bar{f}_{QQ}}{a^2} \left( 3\mathcal{H}^{'} - \mathcal{H}^2 \right) + 6 \frac{\bar{f}^{\,'}_{QQ}}{a^2} \mathcal{H} \biggr) \mathcal{H} \phi^{'} \nonumber \\
&+\frac{1}{3} \bar{f}_Q k^2 \Phi - \frac{1}{3} \left( \bar{f}_Q + 6 \frac{\bar{f}_{QQ}}{a^2} \mathcal{H}^2 \right) k^2 W^{'} - \frac{1}{3} \biggl( 2 \bar{f}_Q + 3 \frac{\bar{f}_{QQ}}{a^2} \left( 5\mathcal{H} - \mathcal{H}^2 \right) + 6 \frac{\bar{f}_{QQ}^{\,'}}{a^2} \mathcal{H} \biggr) \mathcal{H} k^2 W.
\end{align}

This equation also has a coupling between $\delta \rho$ and $\delta p$ introduced by the derivative of $f$ against $T$ ($f_T$). However, this coupling can be broken if we work with radiation since its equation of state is given by $w = p/\rho = 1/3$.\\

As we can see, the presence of derivatives of $f(Q,T)$ with respect to $T$ couples the equations for $\delta \rho$ and $\delta p$. However if we set $\bar{f}_T = 0 = \bar{f}_{TT}$ we recover the equations for $f(Q)$ \cite{jimenez2020cosmology}, as expected. By taking the gradient of $T^0_{\;\;i}$ we can get the equation for $v$
\begin{align}
\label{eqn:velocityPerturbation}
    v &= \frac{1}{(\bar{f}_T + 8\pi)(\bar{\rho}+\bar{p})a^2} \biggl[ 2 \left(\bar{f}_Q+3\frac{\bar{f}_{QQ}}{a^2} (\mathcal{H}'-\mathcal{H}^2)\right) \mathcal{H} \Psi + 12 \frac{\bar{f}_{QQ}}{a^2} \mathcal{H}^2 \phi' \nonumber \\
    &+ 18 \mathcal{H} \frac{\bar{f}_{QQ}}{a^2} (\mathcal{H}'-\mathcal{H}^2) \phi + 2\bar{f}_Q \Phi' - 4k^2 \frac{\bar{f}_{QQ}}{a^2} \mathcal{H}^2 W \biggr]
\end{align}

We can now proceed with the perturbation of the connection equations (\ref{eqn:connectionFieldEquations}). In the case of the zeroth component
\begin{align}
\label{eqn:connectionEquationPerturbation0}
    &k^2 W \bar{f}_{QQ} \mathcal{H}^2 - 6\phi' \bar{f}_{QQ} \mathcal{H}^2 - 6 \Psi \bar{f}_{QQ} \mathcal{H}^3 - 6 \mathcal{H}^2 (\mathcal{H}'-\mathcal{H}^2) \bar{f}_{QQ} W - 3 \mathcal{H} (\mathcal{H}'-\mathcal{H}^2) \bar{f}_{QQ} W' \nonumber \\
    &+ 12 \mathcal{H} (\mathcal{H}'-\mathcal{H}^2) \bar{f}_{QQ} \phi - 9 \mathcal{H} (\mathcal{H}'-\mathcal{H}^2) \bar{f}_{QQ} \phi - 3 \mathcal{H} (\mathcal{H}'-\mathcal{H}^2) \bar{f}'_{QQ} \nonumber - 3\mathcal{H} (\mathcal{H}'-\mathcal{H}^2) \bar{f}_{QQ} \Psi \\
    &= \dfrac{4\pi k_\mu k_\nu \delta H_0^{\;\;\mu\nu}}{k^2}.
\end{align}

Finally, the divergence of the perturbation of the spacial part of the connection equations (\ref{eqn:connectionFieldEquations}) is given by
\begin{align}
    &\mathcal{H}^2 (-18 \bar{f}_{QQ} (\mathcal{H}'+\mathcal{H}^2)\Psi - 2 \bar{f}_{QQ} \mathcal{H} k^2 W + 12 \bar{f}_{QQ} \mathcal{H} \phi' - 6\bar{f}_{QQ}\mathcal{H}\Psi'-6\mathcal{H} \bar{f}'_{QQ} \Psi)  \nonumber \\
    &+ \mathcal{H} (2\bar{f}_{QQ} k^2 W (\mathcal{H}'+\mathcal{H}^2)+6\bar{f}_{QQ} \mathcal{H} \phi'' + \bar{f}_{QQ} \mathcal{H} k^2 W' + 12 \bar{f}_{QQ} (\mathcal{H}'+\mathcal{H}^2)\phi'\nonumber \\
    &+\bar{f}'_{QQ} \mathcal{H} k^2 W - 6\bar{f}'_{QQ} \mathcal{H} \phi') + 18\bar{f}_{QQ} \mathcal{H}^4 \Phi + \frac{3}{2}a^2 \mathcal{H} \bar{f}'_Q \phi + \frac{1}{2} a^2 \mathcal{H} \bar{f}'_Q \Psi + \frac{3}{4} a^2 \bar{f}''_Q \phi \nonumber \\
    &+ \frac{3}{4} a^2 \bar{f}'_Q \phi' - \frac{1}{4} a^2 \bar{f}'_Q k^2 W + \bar{1}{4} a^2 \bar{f}'_Q \Psi' + \frac{3}{2} a^2 \bar{f}'_{QQ} \Psi = -\frac{4\pi i k_i k_\mu k_\nu}{k^2} \delta H_i^{\;\;\mu\nu}.
\end{align}

With the aid of this equations, we will be able to test these models using CMB data. The study of them with such data is beyond the scope of this paper. However, we will give a general qualitative procedure to do it. We need to compute the temperature perturbations in the Fourier space
\begin{equation}
    \Theta(\mathbf{k},\mu,\eta) = \frac{\delta T (\mathbf{k},\mu,\eta)}{T_0(\eta)},
\end{equation}
where $T_0(\eta)$ is the CMB temperature at a given conformal time $\eta$, $\delta T$ the temperature perturbation, $\mathbf{k}$ the Fourier vector, and $\mu = \dfrac{\mathbf{k} \cdot \mathbf{p}}{kp}$ \cite{callin2006calculate,dodelson2020modern} with $\mathbf{p}$ the momentum of the photon. This perturbation can be expanded in multipoles as \cite{callin2006calculate}
\begin{equation}
    \Theta_l = \frac{i^l}{2} \int_{-1}^1 \mathcal{P}_l(\mu) \Theta(\mu) d\mu,
\end{equation}
where $\mathcal{P}_l(\mu)$ are the Legendre polynomials. As a first step, it is needed to transform the scalar perturbation equations to the Newtonian gauge where the scalar perturbations $W$ and $E$ are zero. After this, it is required to solve the Boltzmann equation for the $\Theta$ temperature perturbation which is given in terms of the scalar perturbations \cite{dodelson2020modern}. Then, the CMB power spectrum is \cite{callin2006calculate,dodelson2020modern}
\begin{equation}
    C_l = \int \frac{d^3 k}{(2\pi)^3} P(k) \Theta_l^2(k),
\end{equation}
where $P(k)$ is the primordial power spectrum which can be predicted by inflation \cite{callin2006calculate}. This theoretical value depends on the cosmological parameters \cite{dodelson2020modern,weinberg2008cosmology} $H_0$, $\Omega_m$, etc. Therefore, by computing this power spectrum, we can compare the theoretical predicted values of a given $f(Q,T)$ model with the results from the Planck collaboration \cite{Planck:2018vyg} with a Bayesian analysis. In particular, we can constrain the value of $H_0$ predicted by this class of models. If the value of a given model is consistent with the one of the SH0ES collaboration \cite{Riess:2019cxk}, that model would solve the Hubble constant tension. Therefore, these analyses are important to determine whether $f(Q,T)$ can reduce the Hubble constant tension if their predicted $H_0$ value with CMB data is similar to the one of SH0ES \cite{Riess:2019cxk}.\\

The solutions to the differential equations of the scalar fields and the temperature perturbations have to be done with numerical methods. There are several codes that compute the CMB power spectrum like CLASS \footnote{\url{https://lesgourg.github.io/class_public/class.html}} \cite{lesgourgues2011cosmic,blas2011cosmic}, or CAMB \footnote{\url{https://camb.info/}}. However, these codes work with the background, perturbation, etc. equations of General Relativity. Then, in order to study this class of $f(Q,T)$ models, a new code would be required which takes into consideration the postulates of symmetric teleparallel gravity and the $f(Q,T)$ equations.

\section{Density Contrast Equation}
\label{sec:densityConstrast}

To derive the equation for $\delta = \dfrac{\delta \rho}{\bar{\rho}}$, we need to compute the Continuity and Euler equations. By starting with the former, let us consider the perturbation of the zeroth component of equation (\ref{eqn:energyBalanceEquiation})
\begin{align}
    \label{eqn:continuityEquation}
    &\left[ \bar{f}_T + 8\pi + \frac{1}{2} \bar{f}_T \left( 1 -  \frac{\delta p'}{\delta \rho'} \right) \right] \delta' - \biggl[ 3 \mathcal{H} (\bar{f}_T + 8\pi) \left( \frac{(\bar{f}_T + 8\pi) w - \frac{1}{2} \bar{f}_T (1-c_s^2)}{\bar{f}_T + 8\pi + \frac{1}{2} \bar{f}_T (1-c_s^2)} - \frac{\delta p}{\delta \rho} \right) \nonumber \\
    &+ \frac{3\mathcal{H} (\bar{f}_T + 8\pi) (1+w)}{2\left(\bar{f}_T + 8\pi + \frac{1}{2} \bar{f}_T (1-c_s^2)\right)} \left( \bar{f}_T \left( 1 - \frac{\delta p'}{\delta \rho'} \right) + \bar{f}_{TT} \bar{\rho} \left( 1 - 3 \frac{\delta p}{\delta \rho} \right) (1-3 c_s^2) \right) \nonumber \\
    &+ 3\mathcal{H} \bar{f}_{TT} \bar{\rho} \left( 1 - 3 \frac{\delta p}{\delta \rho} \right) (1+w) \left( \frac{\bar{f}_T (1-c_s^2)}{\bar{f}_T (3-c_s^2) + 16\pi} \right)  \biggr] \delta - (\bar{f}_T + 8\pi ) \biggl[ (1+w) k^2 (v-W) \nonumber \\
    &+ 3\phi' (1+w) \biggr] = \frac{\delta B_0}{\bar{\rho}},
\end{align}
where $\delta B_0$ is the perturbation of the zeroth component of the right-hand side of equation (\ref{eqn:energyBalanceEquiation}). The expression is cumbersome and will be left in appendix (\ref{sec:PertEnergyBalance}). On the other hand, the Euler equation is given by the spatial divergence of equation (\ref{eqn:energyBalanceEquiation})
\begin{align}
\label{eqn:EulerEquation}
    v' + \frac{3\bar{f}_T + 8\pi - c_s^2 (8\pi + 5\bar{f}_T)}{\bar{f}_T + 8\pi + \frac{1}{2} \bar{f}_T (1-c_s^2)} \mathcal{H} v &+ \frac{\delta p}{\bar{\rho} + \bar{p}} + \Psi - \frac{\bar{f}_T}{(\bar{f}_T + 8\pi) \mathcal{H} (\bar{\rho} + \bar{p})} \delta p \nonumber \\
    &= \frac{i k_j \delta B_j}{(\bar{f}_T + 8\pi) (\bar{\rho}+\bar{p})k^2},
\end{align}
where $\delta B_j$ is the perturbation of the j-th component. Since this component is cumbersome, it will be given in appendix (\ref{sec:PertEnergyBalance}). As we can see, we have two coupled equations with $\delta$ and $v$. The general procedure to solve the equation for $\delta$ is to take the conformal time derivative of equation (\ref{eqn:continuityEquation}) and then use equation (\ref{eqn:EulerEquation}) to break the coupling. However, we should not proceed in that way because $\bar{f}_T$ is a function of $\eta$ and hence when taking the derivative with respect to conformal time, the expression will be even bigger. Therefore, if a solution for $\delta$ is required, we should work with the coupled equations and solve them numerically to get the solution for the density contrast $\delta$. We will consider dust and then $w = 0 = c_s^2$, and also $\delta p = 0 = \delta p'$
\begin{align}
\label{eqn:continuityEquationDust}
    \left( \frac{3}{2} \bar{f}_T + 8\pi \right) \delta' &+ \frac{3\mathcal{H}}{3\bar{f}_T + 16\pi} \left[ (\bar{f}_T + 8\pi) \bar{f}_T - (\bar{f}_T + 8\pi) (\bar{f}_T + \bar{f}_{TT} \bar{\rho}) - \bar{f}_{TT} \bar{\rho} \bar{f}_T \right] \delta \nonumber \\
    & - (\bar{f}_T + 8\pi) \left[ k^2(v-W) + 3\phi' \right] = \frac{\delta B_0}{\bar{\rho}}
\end{align}
\begin{align}
\label{eqn:EulerEquationDust}
    v' + \frac{6\bar{f}_T + 16\pi}{3\bar{f}_T + 16\pi} \mathcal{H} v + \Psi = \frac{i k_j \delta B_j}{(\bar{f}_T + 8\pi) \bar{\rho} k^2}.
\end{align}

Therefore, to solve this coupled system, we need to know the evolution of $W$, $\Psi$, $\delta B_0$ and $ik_j \delta B_j$. To compute the evolution of $\Phi$, we can use equation (\ref{eqn:density}). Since we are considering dust, we are considering the period of the universe in which matter dominates, that is, when the perturbations have entered the Hubble radius, in the Sub-Hubble limit $\mathcal{H}<<k$ \cite{alvarenga2013dynamics}. Furthermore, if we neglect the time derivative of the potentials $W, \Phi, \phi, \Psi$, we get
\begin{equation}
    \label{eqn:phi}
    \Phi = \frac{a^2 \delta \bar{\rho}\left(8\pi + \dfrac{3}{2} \bar{f}_T - \bar{f}_{TT} \bar{\rho}\right)}{2 \bar{f}_Q k^2}.
\end{equation}

Let us clarify that neglecting the time derivatives of the potentials and assuming $\mathcal{H}<<k$ does not mean the quasi-static (QS) limit, because that limit is not well defined in these theories \cite{jimenez2020cosmology}. To get the evolution of $\Psi$, we can take equation (\ref{eqn:pressure}) and (\ref{eqn:phi}) in the $\mathcal{H}<<k$ limit and with negligible time derivatives of the potentials
\begin{equation}
    \Psi = \frac{a^2 \delta \bar{\rho} \left( 8\pi - \bar{f}_{TT} \bar{\rho} \right)}{2 \bar{f}_Q k^2}.
\end{equation}

Since the right-hand sides of equations (\ref{eqn:continuityEquationDust}) and (\ref{eqn:EulerEquationDust}) depend on the scalar field $W$, as (\ref{eqn:deltaB0}) and (\ref{eqn:deltaBj}) show us, we need the evolution of this field to solve the equations for $\delta$ and $v$. And the evolution of this field is given in equation (\ref{eqn:velocityPerturbation}) when taking the $\mathcal{H}<<k$ limit and negligible time derivatives. Then the equations are given by
\begin{align}
   & \left( \frac{3}{2} \bar{f}_T + 8\pi \right) \delta' + \frac{3\mathcal{H}}{3\bar{f}_T + 16\pi} \left[ (\bar{f}_T + 8\pi) \bar{f}_T - (\bar{f}_T + 8\pi) (\bar{f}_T + \bar{f}_{TT} \bar{\rho}) - \bar{f}_{TT} \bar{\rho} \bar{f}_T \right] \delta \nonumber \\
   & - \frac{(\bar{f}_T + 8\pi)}{2} \left( k^2  + \frac{(\bar{f}_T+8\pi)\bar{\rho}a^4}{2\bar{f}_{QQ} \mathcal{H}^2} \right) v = 0,
\end{align}
and
\begin{align}
    v' + \frac{6\bar{f}_T + 16\pi}{3\bar{f}_T + 16\pi} \mathcal{H} v + \frac{a^2 \bar{\rho} (8\pi - \bar{f}_{TT} \bar{\rho})}{2\bar{f}_Q k^2} \delta + \frac{1}{2} \left( \frac{\bar{f}'_{QQ}}{\bar{f}_{QQ}} - 3\left(\frac{\mathcal{H}'}{\mathcal{H}}+\mathcal{H}\right) \right) v = 0.
\end{align}

Therefore, we have a system of two equations with two variables $\delta$ and $v$ that can be solved numerically. Note, that they also reduce to the ones for $f(Q)$ \cite{jimenez2020cosmology} if $\bar{f}_T = 0$ because in that case $\delta B_0$ and $i k_j \delta B_j$ are equal to zero and then the extra $v$ terms would be zero in that limit. These results are valid in the coincident gauge. If we needed to know how these equations behave outside this gauge, we would need to perform a gauge transformation because by taking this gauge, we lost the diffeomorphism invariance of the equations. They are also valid whenever $\mathcal{H}<<k$ and the time derivatives of $\Phi,\phi,W, \Psi$ are negligible. These conditions do not mean the QS limit because it has not been well defined in the framework of symmetric teleparallel gravity.

\section{The weak coupling limit}
\label{sec:weakCoupling}

We can study the effects of the coupling between the trace of the stress energy tensor $T$ and the non-metricity scalar $Q$ by considering the weak and strong coupling limits. We will start with the former. If we remember that we are working with $f(Q,T)$ functions of the form $f(Q,T) = f_1(Q) + f_2(T)$, we can study the weak coupling limit when $f_2(T)$ is small. We will perform a Taylor expansion of $f_2(T)$ around $T=0$
\begin{equation}
    \label{eqn:taylorFirstOrderApproximation}
    f_2(T) = f_{2}(0) + \frac{df_2(0)}{dT} T + \frac{1}{2} \frac{d^2 f_2(0)}{dT^2} T^2 + \dots.
\end{equation}

If we take the first order approximation, we can build a small $f_2(T)$ provided that $T$ and $f_2(0)$ are small. We can do this because most $f(Q,T)$ functions that have been studied in the literature \cite{gadbail2021viscous,Agrawal:2021rur,Gadbail:2021kgd,Pati:2021ach,Pradhan:2021dpf,Arora:2021tuh,Godani:2021mld,Arora:2021jik,Bhattacharjee:2021hwm,Zia:2021vhr,Arora:2020iva,Arora:2020tuk,Arora:2020met,najera2021fitting} consider polynomial functions of the form $f_2(T) = \alpha T + \beta T^2$. Therefore, these functions fulfil the criterion of $f_2(0)$ small since $f_2(0)=0$ for this class of functions. The other condition to guarantee the smallness of $f_2(T)$ is that $T$ is small, i.e, that the stress-energy content is small. As we can see, the derivative of $f_2(T)$ against $T$ is just a constant which we will redefine as $\dfrac{df_2(0)}{dT} \equiv \alpha$. In this limit, we can compute the evolution of $\rho$ (\ref{eqn:rhoPrimeBack})
\begin{equation}
    \rho = \rho_0 \left( \frac{a}{a_0} \right)^{-\kappa},
\end{equation}
\textbf{with}
\begin{equation}
    \kappa = \dfrac{3(1+w)(\alpha+8\pi)}{8\pi+\frac{1}{2}\alpha(3-w)},
\end{equation}
and $\rho_0 = \rho(a_0)$. We have also assumed a constant equation of state $w$. Hence, if we break the coupling, i.e, we set $\alpha=0$, we recover the standard equation for the density evolution. Moreover, the coupling causes changes in the density evolution with the Hubble flow in the weak coupling limit. If $\alpha>0$, the density would be smaller than in the standard evolution $\rho = \rho_0 \left( \dfrac{a}{a_0} \right)^{-3}$ and it would be bigger if $\alpha<0$. We can now see what happens with the vector perturbations. In this weak coupling limit, the $v_i$ vector perturbation is given by
\begin{equation}
\label{eqn:velocityVectorWeakLimit}
    v_i = -\frac{\bar{f}_Q k^2}{2(\alpha+8\pi)\bar{\rho}(1+w)a^2} (E'_i-W_i),
\end{equation}
as we can see, the modulus of this vector decreases if $\alpha>0$ and it increases if $\alpha<0$. For scalar perturbations, we start with equations (\ref{eqn:density})
\begin{align}
\label{eqn:densityWeak}
& a^2 \delta \rho \left( 8\pi + \frac{3}{2} \alpha \right) - \frac{1}{2}a^2 \alpha \delta p \nonumber \\
&= 6\left( \bar{f}_Q + 12 \frac{\bar{f}_{QQ}}{a^2} \mathcal{H}^2 \right) \mathcal{H} \left(\mathcal{H}\Psi + {\Phi}^{'}\right) + 2\bar{f}_Q k^2 \Phi -2\left( \bar{f}_Q + 3 \frac{\bar{f}_{QQ}}{a^2} \left( {\mathcal{H}}^{'} + \mathcal{H}^2 \right) \right) \mathcal{H} k^2 W,
\end{align}
and (\ref{eqn:pressure})
\begin{align}
\label{eqn:pressureWeak}
& \frac{1}{4} a^2 \alpha \delta \rho - a^2 \left( 4\pi + \dfrac{3}{4} \alpha \right) \delta p = \left( \bar{f}_Q + 12 \frac{\bar{f}_{QQ}}{a^2} \mathcal{H}^2 \right) \left( \mathcal{H} {\Psi}^{'} + {\phi}^{''} \right) + \biggl( \bar{f}_Q \left( {\mathcal{H}}^{'} + 2\mathcal{H}^2 - \dfrac{1}{3} k^2 \right) \nonumber \\
&+ 12 \frac{\bar{f}_{QQ}}{a^2} \mathcal{H}^2 \left(4 {\mathcal{H}}^{'} - \mathcal{H}^2 \right) + 12 \frac{{\bar{f}_{QQ}}^{\,'}}{a^2} \mathcal{H}^3 \biggr) \Psi + 2 \biggl( \bar{f}_Q + 6 \frac{\bar{f}_{QQ}}{a^2} \left( 3\mathcal{H}^{'} - \mathcal{H}^2 \right) + 6 \frac{\bar{f}^{\,'}_{QQ}}{a^2} \mathcal{H} \biggr) \mathcal{H} \phi^{'} \nonumber \\
&+\frac{1}{3} \bar{f}_Q k^2 \Phi - \frac{1}{3} \left( \bar{f}_Q + 6 \frac{\bar{f}_{QQ}}{a^2} \mathcal{H}^2 \right) k^2 W^{'} - \frac{1}{3} \biggl( 2 \bar{f}_Q + 3 \frac{\bar{f}_{QQ}}{a^2} \left( 5\mathcal{H} - \mathcal{H}^2 \right) + 6 \frac{\bar{f}_{QQ}^{\,'}}{a^2} \mathcal{H} \biggr) \mathcal{H} k^2 W.
\end{align}

Hence, even in the weak coupling limit, the coupling between $\delta \rho$ and $\delta p$ exists. Finally, let us analyse the behaviour of equation (\ref{eqn:velocityPerturbation})
\begin{align}
    v &= \frac{1}{(\alpha + 8\pi)(\bar{\rho}+\bar{p})a^2} \biggl[ 2 \left(\bar{f}_Q+3\frac{\bar{f}_{QQ}}{a^2} (\mathcal{H}'-\mathcal{H}^2)\right) \mathcal{H} \Psi + 12 \frac{\bar{f}_{QQ}}{a^2} \mathcal{H}^2 \phi' \nonumber \\
    &+ 18 \mathcal{H} \frac{\bar{f}_{QQ}}{a^2} (\mathcal{H}'-\mathcal{H}^2) \phi + 2\bar{f}_Q \Phi' - 4k^2 \frac{\bar{f}_{QQ}}{a^2} \mathcal{H}^2 W \biggr],
\end{align}
and then, the weak coupling has the effect to decrease the value of $v$ if $\alpha>0$ and to increase it if $\alpha<0$. Then, it has a similar effect to the one in equation (\ref{eqn:velocityVectorWeakLimit}). To finish with the weak coupling limit, we need to study the density contrast equations (for the case w=0). By setting $f_T = \alpha$, we get
\begin{align}
\label{eqn:densityContrast1Weak}
    \left( 1 + \frac{3\alpha}{16\pi}\right) \delta' - \left(\frac{1}{2} + \frac{\alpha}{16\pi}\right) \left( k^2  + \frac{(\alpha+8\pi)\bar{\rho}a^4}{2\bar{f}_{QQ} \mathcal{H}^2} \right) v = 0,
\end{align}
and
\begin{align}
    v' + \frac{6\alpha + 16\pi}{3\alpha + 16\pi} \mathcal{H} v + \frac{4\pi a^2 \bar{\rho}}{\bar{f}_Q k^2} \delta + \frac{1}{2} \left( \frac{\bar{f}'_{QQ}}{\bar{f}_{QQ}} - 3\left(\frac{\mathcal{H}'}{\mathcal{H}}+\mathcal{H}\right) \right) v = 0.
\end{align}

As we can see, in this weak coupling limit, the term in equation (\ref{eqn:densityContrast1Weak}) proportional to $\delta$ disappears. Also, if $\alpha <<8\pi$, we recover the $f(Q)$ behaviour. Moreover, even though we are working in the weak coupling limit, the coupling between $\delta \rho$ and $\delta p$ persists in equations (\ref{eqn:densityWeak}) and (\ref{eqn:pressureWeak}).

\section{The Strong Coupling Limit}
\label{sec:strongCoupling}

This limit is much harder to study because in this case we need $f_2(T)$ to be big. However, we can simplify this problem since most $f(Q,T)$ models in the literature consider functions of the form $f_2(T) = \alpha T + \beta T^2$ \cite{gadbail2021viscous,Agrawal:2021rur,Gadbail:2021kgd,Pati:2021ach,Pradhan:2021dpf,Arora:2021tuh,Godani:2021mld,Arora:2021jik,Bhattacharjee:2021hwm,Zia:2021vhr,Arora:2020iva,Arora:2020tuk,Arora:2020met,najera2021fitting}. Hence, with this specific function of $f_2(T)$, we can guarantee that it is big provided that $T$ is big. Its derivative is given by $f_{2T}(T) = \alpha + 2 \beta T$, which is also big because of the dependence on $T$, unless $\beta=0$. In this case, we cannot solve equation (\ref{eqn:rhoPrimeBack}) in an analytic way as we did in the weak coupling limit. And even though the function $f_T$ is not simply a constant, we can still analyse the behaviour of the perturbation equations in a qualitative way. The vector perturbations are given by
\begin{equation}
\label{eqn:velocityPerturbationStrong}
    v_i \approx - \frac{\bar{f}_Q k^2}{2(\alpha + 2\beta \bar{\rho}(3w-1)) \bar{\rho} a^2 (1+w)} \left( E'_i - W_i \right).
\end{equation}

Since $T$ is big, in this strong coupling limit, the modulus of the vector perturbations are small and can be negligible in the limit $\bar{f}_Q <<\bar{f}_T$. However, in the especial case $\beta=0$, the behaviour is identical to the one of the weak coupling limit. We can now study the effect on the scalar perturbations. We can start with equations (\ref{eqn:density})
\begin{align}
& a^2 \delta \rho \left( \frac{3}{2}\alpha + \beta \bar{\rho} (7w-5) \right) + a^2 \delta p \left( -\frac{1}{2} \alpha + \beta \bar{\rho} (7+3w) \right) \nonumber \\
&= 6\left( 8\pi + \bar{f}_Q + 12 \frac{\bar{f}_{QQ}}{a^2} \mathcal{H}^2 \right) \mathcal{H} \left(\mathcal{H}\Psi + {\Phi}^{'}\right) + 2\bar{f}_Q k^2 \Phi -2\left( \bar{f}_Q + 3 \frac{\bar{f}_{QQ}}{a^2} \left( {\mathcal{H}}^{'} + \mathcal{H}^2 \right) \right) \mathcal{H} k^2 W,
\end{align}
and
\begin{align}
& \frac{1}{4} a^2 \delta \rho \left( \alpha + 2\beta \bar{\rho}(3w-1) \right) - a^2 \delta p\left( 4\pi + \dfrac{3}{4} (\alpha + 2\beta\bar{\rho} (3w-1)) \right) = \left( \bar{f}_Q + 12 \frac{\bar{f}_{QQ}}{a^2} \mathcal{H}^2 \right) \nonumber \\
&\times \left( \mathcal{H} {\Psi}^{'} + {\phi}^{''} \right) + \biggl( \bar{f}_Q \left( {\mathcal{H}}^{'} + 2\mathcal{H}^2 - \dfrac{1}{3} k^2 \right) + 12 \frac{\bar{f}_{QQ}}{a^2} \mathcal{H}^2 \left(4 {\mathcal{H}}^{'} - \mathcal{H}^2 \right) + 12 \frac{{\bar{f}_{QQ}}^{\,'}}{a^2} \mathcal{H}^3 \biggr) \Psi \nonumber \\
&+ 2 \biggl( \bar{f}_Q + 6 \frac{\bar{f}_{QQ}}{a^2} \left( 3\mathcal{H}^{'} - \mathcal{H}^2 \right) + 6 \frac{\bar{f}^{\,'}_{QQ}}{a^2} \mathcal{H} \biggr) \mathcal{H} \phi^{'} +\frac{1}{3} \bar{f}_Q k^2 \Phi - \frac{1}{3} \left( \bar{f}_Q + 6 \frac{\bar{f}_{QQ}}{a^2} \mathcal{H}^2 \right) k^2 W^{'} \nonumber \\ 
&- \frac{1}{3} \biggl( 2 \bar{f}_Q + 3 \frac{\bar{f}_{QQ}}{a^2} \left( 5\mathcal{H} - \mathcal{H}^2 \right) + 6 \frac{\bar{f}_{QQ}^{\,'}}{a^2} \mathcal{H} \biggr) \mathcal{H} k^2 W,
\end{align}
as we can see, in this limit, the coupling between $\delta \rho$ and $\delta p$ is also strong unless $\beta = 0$ which is a possibility since $f_2(T) = \alpha T$ is also a strong coupling limit function. Now, we continue with equation (\ref{eqn:velocityPerturbation})
\begin{align}
    v &= \frac{1}{(\alpha + 2\beta \bar{\rho}(3w-1) + 8\pi)(\bar{\rho}+\bar{p})a^2} \biggl[ 2 \left(\bar{f}_Q+3\frac{\bar{f}_{QQ}}{a^2} (\mathcal{H}'-\mathcal{H}^2)\right) \mathcal{H} \Psi + 12 \frac{\bar{f}_{QQ}}{a^2} \mathcal{H}^2 \phi' \nonumber \\
    &+ 18 \mathcal{H} \frac{\bar{f}_{QQ}}{a^2} (\mathcal{H}'-\mathcal{H}^2) \phi + 2\bar{f}_Q \Phi' - 4k^2 \frac{\bar{f}_{QQ}}{a^2} \mathcal{H}^2 W \biggr],
\end{align}
which has a similar behaviour to equation (\ref{eqn:velocityPerturbationStrong}). When $\beta \neq 0$, the value of this scalar perturbation is significantly reduced and can tend to 0 in the $\bar{f}_Q <<\bar{\rho}$ and $\bar{f}_{QQ} <<\bar{\rho}$ limits. However, when $\beta=0$, this perturbations behaves as the one corresponding to the weak coupling limit. Finally, the contrast equations for the $w=0$ case are given by
\begin{align}
   \left( \frac{3}{2} \alpha - 3\beta \bar{\rho} + 8\pi \right) \delta' - &\frac{12\mathcal{H}\beta\bar{\rho}}{3\alpha - 6\beta \bar{\rho} + 16\pi} \left[ \alpha - 2\beta\bar{\rho} +4\pi \right] \delta \nonumber \\
   & - \frac{(\alpha - 2\beta\bar{\rho} + 8\pi)}{2} \left( k^2  + \frac{(\alpha - 2\beta \bar{\rho}+8\pi)\bar{\rho}a^4}{2\bar{f}_{QQ} \mathcal{H}^2} \right) v = 0,
\end{align}
and
\begin{align}
    v' + \frac{6\alpha - 12\beta\bar{\rho} + 16\pi}{3\alpha - 6\beta \bar{\rho} + 16\pi} \mathcal{H} v + \frac{a^2 \bar{\rho} (4\pi - \beta \bar{\rho})}{\bar{f}_Q k^2} \delta + \frac{1}{2} \left( \frac{\bar{f}'_{QQ}}{\bar{f}_{QQ}} - 3\left(\frac{\mathcal{H}'}{\mathcal{H}}+\mathcal{H}\right) \right) v = 0.
\end{align}

If $\beta\neq0$, the density contrast is heavily driven by the background density. Moreover, if $\beta=0$, we recover the weak coupling limit.\\

As we can see, in the strong coupling limit, for a specific case where $f_2(T) = \alpha T + \beta T^2$ (a case broadly studied in the literature), the perturbation equations are heavily driven by $\bar{\rho}$, the background density, when $\beta\neq0$. However, when $\beta = 0$, we recover the weak coupling limit results even though we are working with non-minimally coupling $f_2(T)$ functions. This is an interesting and important result since most $f_2(T)$ functions considered in the literature are of the form $f_2(T) = \alpha T$. Therefore, for this class of models, the behaviour of their equations is weak coupling limit like.

\section{Conclusions}
\label{sec:conclusions}

 In the present paper, we have developed the linear theory of perturbations in $f(Q,T)$ theories with $Q$ the non-metricity scalar and $T$ the trace of the stress-energy tensor, which are an extension of symmetric teleparallel gravity, with an specific form $f(Q,T) = f_1(Q) + f_2(T)$ that has been studied extensively in the literature. By taking this ansatz for the $f(Q,T)$ functions, we got equations consistent with $f(Q)$ gravity \cite{jimenez2020cosmology} at the limit $\bar{f}_T = 0$. However, outside of this limit, the coupling of $Q$ and $T$ in the Lagrangian induces a coupling between the perturbation of the density and the pressure in the case of scalar perturbation equations. This coupling disappears for the $\delta p$ equation when considering radiation (a fluid with equation of state $w=1/3$). In the weak coupling limit (when $f_2(T)$ is small), this coupling still exists unless $d f_2(0)/dT << 8\pi$. On the other hand, considering a function $f_2(T) = \alpha T + \beta T^2$ which has been widely studied in the literature, and the strong coupling limit ($T$ big), the perturbative equations are heavily driven by the stress-energy terms when $\beta \neq 0$. However, when $\beta=0$, the perturbative equations are identical to the ones of the weak coupling limit. Therefore, models of the form $f(Q,T) = f_1(Q) + \alpha T$ have a weak coupling like behaviour even when considering non-minimally coupling between $Q$ and $T$ in the Lagrangian. Also, the presence of $T$ in the Lagrangian breaks the equality to zero of the connection equations by introducing the hypermomentum. Its presence also breaks the stress-energy conservation by inducing stress-energy transfer between geometry and matter and particle creation/annihilation \cite{xu2019f,wu2018palatini}. \\

By computing the perturbations of this class of theories, we will enable future early universe studies such as early universe constraints of cosmological parameters that will shed light on whether these theories can challenge the concordance $\Lambda$CDM model at a perturbative and early universe perspectives apart from the background one where they have already been found to challenge $\Lambda$CDM \cite{najera2021fitting}. In addition to this, with the aid of the tensor perturbative equations, future  studies of $f(Q,T)$ with the aid of standard sirens will be possible. Since in modified gravity theories, the standard luminosity distance and gravitational waves luminosity distance differ, standard sirens will constitute a valuable tool to constrain deviations from GR \cite{belgacem2018gravitational}.\\

We have also provided the overdensity equation for dust with $w=0$ with two coupled differential equations between $\delta$ and $v$ in the $\mathcal{H}<<k$ limit with negligible time derivative of the scalar potentials and the coincident gauge. A future work will test, with the aid of this differential equations, if $f(Q,T)$ can be considered an alternative to dark matter. These study should be done carefully because the equations that we derived are valid in the $\mathcal{H}<<k$ limit and with negligible time derivatives of the scalar potentials, which does not mean the quasi-static (QS) limit due to the fact that this limit is not well defined in the symmetric teleparallel gravity framework. Furthermore, the perturbation equations might enable future $f(Q,T)$ studies with CMB and standard siren data such as Planck \cite{Planck:2018vyg}, Atacama Cosmology Telescope \cite{Mallaby-Kay:2021tuk}, or LISA \cite{belgacem2019testing}. These analyses will help to see whether $f(Q,T)$ gravity can reduce the Hubble Constant tension and therefore constitute an alternative to $\Lambda$CDM. These studies will be reported elsewhere. 

\appendix
\section{Perturbation of the Right-Hand Side of the Energy Balance Equation}
\label{sec:PertEnergyBalance}

In this appendix, we will give the perturbation results of the right-hand side of the energy balance equation. Let us consider the perturbation of the zeroth component of the right-hand side of equation (\ref{eqn:fieldEquationsIndexRaised})

\begin{align}
    &\delta B_0 = \frac{k^2}{a^2}\biggl[2 \Phi \bar{f}'_Q - \frac{3}{2} \phi \bar{f}'_Q - \frac{1}{2} \Psi \bar{f}'_Q - \frac{1}{2}W \bar{f}''_Q - \frac{1}{2} \bar{f}'_Q W' + \mathcal{H} \left( 8 \bar{f}_Q \phi - 3 \bar{f}'_Q W \right) \nonumber \\
    &+ \frac{\mathcal{H}^2}{a^2} \bar{f}_{QQ} \left( 2k^2 W - 12 \phi' \right) - 8 \mathcal{H}^2 \bar{f}_Q W - \frac{12}{a^2} \mathcal{H}^3 \Psi \bar{f}_{QQ}\biggr]. 
\end{align}

We now compute the divergence of the spacial components of equation (\ref{eqn:fieldEquationsIndexRaised})
\begin{align}
    i k_j \delta B_j &= \frac{k^2}{a^2} \biggl[ \frac{84}{a^2} \bar{f}_{QQ} \mathcal{H}^4 \Psi + \frac{1}{a^2} \mathcal{H} \biggl( 4 \bar{f}_{QQ} k^2 W (\mathcal{H}'+\mathcal{H}^2) - 12 \bar{f}_{QQ} \mathcal{H} \phi'' + 2 \bar{f}_{QQ} \mathcal{H} k^2 W' \nonumber \\
    &- 24 \bar{f}_{QQ} (\mathcal{H}'+\mathcal{H}^2) \phi' + 2 \mathcal{H} \bar{f}'_{QQ} k^2 W - 12 \mathcal{H} \bar{f}'_{QQ} \phi' \biggr) - 3\bar{f}_Q (\mathcal{H}'+\mathcal{H}^2) \phi  \nonumber \\
    &+ 2 \bar{f}_Q (\mathcal{H}'+\mathcal{H}^2) \Psi  - 3 \bar{f}_Q (\mathcal{H}'+\mathcal{H}^2) \Phi + 8\bar{f}_Q \mathcal{H} \Phi'  + 3 \mathcal{H} \bar{f}'_{Q} \phi + 5\mathcal{H} \bar{f}'_Q \Psi + \frac{3}{2} \bar{f}''_{Q} \phi  \nonumber \\
    &+ \frac{1}{2} \bar{f}'_{Q} \Psi - \frac{1}{2} k^2 \bar{f}'_{Q} W +\frac{3}{2} \bar{f}_Q \phi' + \frac{1}{2} \bar{f}'_Q \Psi' - \frac{\mathcal{H}^2}{a^2} \biggl( 36 \bar{f}_{QQ} (\mathcal{H}'+\mathcal{H}^2) \Psi + 12 \bar{f}_{QQ} \mathcal{H} k^2 W  \nonumber \\
    &- 24 \bar{f}_{QQ} \mathcal{H} \phi' + 12  \bar{f}_{QQ} \mathcal{H} \Psi' + 12 \bar{f}'_{QQ} \mathcal{H} \Psi \biggr) + 18 \mathcal{H}^2 \bar{f}_Q \phi + 10 \mathcal{H}^2 \bar{f}_Q \Psi \biggr].
\end{align}

Since this results have too much terms, it would be difficult to solve equations (\ref{eqn:continuityEquationDust}) and (\ref{eqn:EulerEquationDust}), however we can consider the evolution in the $\mathcal{H}<<k$ limit and with negligible time derivatives of the scalar potentials. With this conditions, these results reduce to
\begin{equation}
\label{eqn:deltaB0}
    \delta B_0 = \frac{2k^4 \mathcal{H}^2 \bar{f}_{QQ} W}{a^4},
\end{equation}
and
\begin{equation}
\label{eqn:deltaBj}
    i k_j \delta B_j = \frac{2k^4 \mathcal{H}  W }{a^4} \left( \mathcal{H} \bar{f}'_{QQ} - 3(\mathcal{H}'+\mathcal{H}^2) \bar{f}_{QQ} \right).
\end{equation}

Notice that both results depend on $W$. Hence, we need to get the evolution of this scalar field. We can get it by taking the $\mathcal{H}<<k$ limit with negligible time derivatives of the potentials in equation (\ref{eqn:velocityPerturbation})  
\begin{equation}
\label{eqn:velocitySmallScales}
    v = -\frac{4k^2 \bar{f}_{QQ} \mathcal{H}^2 W}{(\bar{f}_T+8\pi)\bar{\rho}a^4}.
\end{equation}

Substituting these results in equations (\ref{eqn:deltaB0}) and (\ref{eqn:deltaBj}) give
\begin{equation}
    \delta B_0 = -\frac{(\bar{f}_T+8\pi) \bar{\rho} k^2}{2} v,
\end{equation}
and
\begin{equation}
    ik_j\delta B_j = - \frac{k^2 (\bar{f}_T+8\pi)\bar{\rho}}{2\mathcal{H}} \left( \mathcal{H} \frac{\bar{f}'_{QQ}}{\bar{f}_{QQ}} - 3(\mathcal{H'} + \mathcal{H}^2) \right) v.
\end{equation}

In this way the perturbations $\delta B_0$ and $i k_j \delta B_j$ are written in terms of $v$. This enables us to solve the equations for $\delta$ (\ref{eqn:continuityEquationDust}) and $v$ (\ref{eqn:EulerEquationDust}). 

\acknowledgments

The perturbation equation computations were done with the Python \texttt{Pytearcat} package. 


\bibliographystyle{JHEP}
\bibliography{references}



\end{document}